\documentclass[11pt,showpacs]{revtex4}
\usepackage{graphicx,amsmath,amssymb,amsfonts,latexsym,color,dcolumn}

\def\be{\begin{equation}}
\def\ee{\end{equation}}

\def\lb{\left[}
\def\rb{\right]}
\def\om{\omega}

\begin{document}

\title{Renormalization in theories with modified dispersion relations: weak gravitational fields}
\author{D. L\'opez Nacir \footnote{dnacir@df.uba.ar}}
\author{F. D. Mazzitelli \footnote{fmazzi@df.uba.ar}}
\affiliation{Departamento de F\'\i sica {\it Juan Jos\'e
Giambiagi}, Facultad de Ciencias Exactas y Naturales, UBA, Ciudad
Universitaria, Pabell\' on I, 1428 Buenos Aires, Argentina}

\begin{abstract}
We consider a free quantum scalar field satisfying modified
dispersion relations in curved spacetimes,  within the framework
of Einstein-Aether theory. Using a power counting analysis, we
study the divergences in the adiabatic expansion of
$\langle\phi^2\rangle$ and $\langle T_{\mu\nu} \rangle$, working
in the weak field approximation. We show that for dispersion
relations containing up to $2s$ powers of the spatial momentum,
the subtraction necessary to renormalize these two quantities on
general backgrounds depends on $s$ in a qualitatively different
way: while $\langle\phi^2\rangle$ becomes convergent for a
sufficiently large value of $s$, the number of divergent terms in
the adiabatic expansion of $\langle T_{\mu\nu}\rangle$ increases
with $s$. This property was not apparent in previous results for 
spatially homogeneous backgrounds.
\end{abstract}


\maketitle

There are theoretical and phenomenological motivations for studying
quantum fields with modified dispersion relations (MDR) in the
effective field theory framework for semiclassical gravity
\cite{amelino,unruh,kosteleky,eatheory,loop,grav,Jacobson,bran1}.
For example, in some inflationary  models the expansion of the
universe can last sufficiently long that length scales of
cosmological interest today could have been sub-Planckian (or
smaller than a critical length for which new physics could show
up) at the beginning of inflation. In such models, the
inflationary predictions could be affected by such unknown
physics, giving rise to so-called trans-Planckian effects
\cite{bran1}. A similar argument in the context of black hole
physics has motivated the analysis of the robustness of the
Hawking effect \cite{unruh} under the modification of the
dispersion relation of the quantum fields. It has also been argued
that trans-Planckian effects could be relevant in the
astrophysical context, as for example in the physics of ultra high
energy cosmic rays \cite{amelino}.

In some models, the inclusion of fields satisfying MDR locally
breaks  Lorentz invariance. Nevertheless, one can include MDR
while preserving a generally covariant metric theory for gravity
by working in the framework of the Einstein-Aether theory
\cite{eatheory}. In this theory, the general covariance is
preserved by introducing a dynamical vector field $u^{\mu}$ called
the aether field, which is constrained by means of a Lagrange
multiplier to take a non-zero timelike value, $u^{\mu}u_{\mu}=-1$.

In the semiclassical approximation,  a study of the consequences of
changing the dispersion relation should involve a proper treatment
of the divergences that appear in the expectation values of
observables associated to the quantum fields. For example, a
careful evaluation of the expectation value of the stress tensor
is necessary to evaluate the backreaction of the quantum fields on
the background metric, since this expectation value is the source
term in the semiclassical equations for the metric.

The renormalization procedure for quantum fields satisfying the
standard dispersion relations in curved backgrounds is well
established \cite{wald,fulling,birrell}. Indeed, there are well
known covariant methods of renormalization that can be implemented
in principle in any spacetime metric. When applied to the
expectation value of the square of the field $\langle
\phi^2\rangle $, or to the mean value of the stress tensor
$\langle T_{\mu\nu} \rangle$, one can obtain the associated
renormalized quantities by making the subtractions:
 \begin{subequations}\label{renor} \begin{align} \label{phiren}
\langle \phi^2 \rangle_{\mathrm{ren}}&=\langle \phi^2
\rangle-\langle \phi^2
\rangle^{(0)}...-\langle \phi^2
\rangle^{(2 i_{max})},\\
\langle T_{\mu\nu} \rangle_{\mathrm{ren}}&=\langle T_{\mu\nu}
\rangle-\langle T_{\mu\nu} \rangle^{(0)}...-\langle T_{\mu\nu}
\rangle^{(2 j_{max})},\label{tren}\end{align}\end{subequations}
where a superscript $2l$ denotes the terms of adiabatic order $2l$
of the corresponding expectation value (i.e., the terms containing
$2l$ derivatives of the metric). For the usual dispersion
relation, in $n$ dimensions the subtraction involves up to the
adiabatic orders $2i_{max}^u=2\, \mathrm{int}(n/2-1)$ for $\langle
\phi^2 \rangle$ and $2j_{max}^u=2\, \mathrm{int} (n/2)$ for the
stress tensor, where $\mathrm{int}(x)$ is the integer part of $x$
(the superscript $u$ stands for {\it usual}).

In the case of MDR, the renormalization procedure have been
investigated in some particular cases \cite{RW12,Bianchi,rinaldi}.
For scalar fields propagating in an $n$-dimensional  spatially
flat Robertson-Walker (RW) spacetime, in Ref. \cite{RW12} we have
considered the extension of the adiabatic subtraction scheme based
on a WKB expansion of the field modes \cite{fpb74} to the case of
generalized dispersion relation.  There, the WKB expansion of the
stress tensor was computed up to the fourth adiabatic order.  With
the use of dimensional regularization, it was shown that these
adiabatic orders can be absorbed into a redefinition of the
gravitational bare constants of the theory (i.e, the cosmological
constant, the Newton constant and the coupling constants
associated to terms quadratic in the curvature in the
gravitational action), which correspond to the counterterms also
required for the usual dispersion relation. However, this simple
result is a peculiarity due to the symmetries of the spatially
flat RW metric. Indeed, in Ref.\cite{Bianchi} we have followed the
same approach for the case of four-dimensional Bianchi type I
metrics. Restricting the calculation up to the second adiabatic
order, we have shown that new counterterms do appear, which
involve the aether field in addition to the metric. For instance,
a term proportional to $(\nabla_{\mu}u^{\mu})^2$ in the aether
Lagrangian is needed in order to absorb the divergences in
$\langle T_{\mu\nu}\rangle^{(2)}$ (in addition to the usual
Einstein-Hilbert action).  The point is that in a spatially flat
RW background these new counterterms are indistinguishable from
the usual ones. Concretely, once evaluated in this background, the
stress tensor obtained from the variation of the most general
action for the aether field containing two derivatives, turns out
to be proportional to the Einstein tensor. As there are strong
constraints on the parameters associated to terms containing two
derivatives of the vector field \cite{Jacobson}, the new
counterterms of second adiabatic order should be carefully chosen
to make the theory consistent with observation \cite{Bianchi}.

In both spatially flat Robertson-Walker and Bianchi type I
backgrounds, given a MDR such that the frequency behaves as
$\om\sim |\vec{k}|^s$ for large values of $|\vec{k}|$, one can
show that divergences appear up to
\cite{RW12,Bianchi}\be\label{jmaxhomog}  2i_{max}^h=2\,
\mathrm{int}\left(\frac{n-1}{2 s}-\frac{1}{2}\right),\,\,\,
2j_{max}^h=2\, \mathrm{int}\left(\frac{1}{2}+\frac{n-1}{2
s}\right),\ee where the superscript $h$ stands for (spatially) {\it
homogeneous}. Note that the correct values for the usual
dispersion relation are recovered for $s=1$. Eq.(\ref{jmaxhomog})
suggests that the higher the power $s$ of $|\vec{k}|$ the milder
would be the divergences. For example, in four dimensions, this
result indicates that if $s\geq 4$ the divergences of the stress
tensor are contained only in the zeroth adiabatic order. However,
as will be shown below, this is valid only for the spatially
homogeneous backgrounds considered so far.

In this letter we will consider weak (but otherwise general)
background fields, and use a power counting analysis to
investigate up to which adiabatic order the subtractions in
Eq.(\ref{renor}) have to be carried out in order to get finite,
physically meaningful expectation values. In the case of
interacting quantum fields in Minkowski spacetime, higher spatial
derivatives improve the UV behavior of Feynman diagrams (see for
instance  \cite{Anselmi}). Here, we will show that while such
improvement also occurs for $\langle \phi^2 \rangle$, this is not
the case for $\langle T_{\mu\nu}\rangle$.

By adopting the semiclassical approximation we consider a massive
quantum scalar field $\phi$ propagating in a classical curved
spacetime and coupled to a classical aether. The classical action
for the scalar field is given by \cite{lemoine,convention}: \be
\label{accionphi} S_{\phi}=\int d^n x
\sqrt{-g}(\mathcal{L}_{\phi}+\mathcal{L}_{cor}),\ee where $n$ is
the spacetime dimension, $g=det(g_{\mu\nu})$, $\mathcal{L}_{\phi}$
is the standard Lagrangian of a free scalar field \be
\mathcal{L}_{\phi}=-\frac{1}{2}\lb g^{\mu
\nu}\partial_{\mu}\phi\partial_{\nu}\phi+(m^2+\xi R)\phi^2\rb,\ee
with $R$ the Ricci scalar, $\mathcal{L}_{cor}$ is the corrective
lagrangian that gives rise to a generalized dispersion relation
\be \mathcal{L}_{cor}=-\sum_{q,p\leq q} b_{qp}
(\mathcal{D}^{2q}\phi)(\mathcal{D}^{2p}\phi),\ee with
$\mathcal{D}^{2}\phi\equiv\perp_{\mu}^{\lambda}\nabla_{\lambda}(\perp_{\gamma}^{\mu}\nabla^{\gamma}\phi$),
where $\perp_{\mu\nu}\equiv g_{\mu\nu}+ u_{\mu} u_{\nu}$ and
$\nabla_{\mu}$ is the derivative operator associated with
$g_{\mu\nu}$.

After varying the action (\ref{accionphi}) with respect to $\phi$,
one can write the equation for the associated Green's function
$G(x,x')$ as \be\label{eqGreen} \left[\Box -(m^2+\xi R)
-\sum_{q,p\leq q}b_{q p}\left(H^q \mathcal{D}^{2p}+H^p
\mathcal{D}^{2q}\right)\right] G(x,x')=-\frac{1}{\sqrt{|g|}}\delta
(x-x'), \ee where $H$ is an operator given by \be
H=\mathcal{D}^{2}+\nabla_{\alpha}a^{\alpha}+2
a^{\alpha}\nabla_{\alpha}, \ee with $a^{\alpha}\equiv
u^{\mu}\nabla_{\mu}u^{\alpha}$.

In the weak field approximation, we consider the linearized
Eq.(\ref{eqGreen}) about the flat background solution with
Minkowski metric $\eta_{\mu\nu}$ and constant aether
$\underline{u}_{\mu}$ \cite{eawaves}. Adopting Minkowski
coordinates $(x^0,...,x^{n-1})$ such that
$\underline{u}_{\mu}=\delta_{\mu}^0$, the fields can be expanded
as
\begin{subequations}\begin{align}
g_{\mu\nu}&=\eta_{\mu\nu}+h_{\mu\nu}\,,\,g^{\mu\nu}=\eta^{\mu\nu}-h^{\mu\nu}, \\
u_{\mu}&=\delta_{\mu}^0+v_{\mu}\,,\,u^{\mu}=-\delta_{0}^{\mu}-h^{0\mu}+v^{\mu}.\end{align}
\end{subequations}
Here and in what follows, we keep only linear terms in
$h_{\mu\nu}$ and $v_{\mu}$. The constraint $u_{\mu}u^{\mu}=-1$
yields $v^0=h^{00}/2$. After choosing the Lorentz gauge
\begin{subequations}
\begin{align}
&\partial^{\mu}\overline{h}_{\mu\nu}=0,\\
&\overline{h}_{\mu\nu}=h_{\mu\nu}-\frac{1}{2}\eta^{\mu\nu}h,
\end{align}
\end{subequations}with $h=h^{\rho}_{\rho}$, we obtain
\begin{subequations}\label{oplin}\begin{align}
&\sqrt{|g|}\Box=(\eta^{\mu\nu}-\overline{h}^{\mu\nu})\partial_{\mu}\partial_{\nu},\\
&\sqrt{|g|}\sum_{q,p\leq q}b_{q p}\left(H^q \mathcal{D}^{2p}+H^p
\mathcal{D}^{2q}\right)=2\sum_{q,p\leq q}b_{q
p}\triangle^{q+p}+h\sum_{q,p\leq q}b_{q
p}\triangle^{q+p}+\mathcal{K},\label{a}
\end{align}
\end{subequations} where $\triangle$ is the Laplacian  and  $\mathcal{K}$ is an
 operator of first order in the perturbation fields $h_{\mu\nu}$ and $v_{\mu}$. The explicit
 expression for the operator  $\mathcal{K}$  can be easily found in the particular case in
 which the perturbation fields depend only on the time coordinate $x^0$. In the
general case, although more complex, it can also be worked out.
 However, the expression in Eq.(\ref{a}) will be enough  for our present purposes.

In order to solve Eq.(\ref{eqGreen}), we
first split $G=G^0+G^1$,  where the superscripts refer to the order
in  $h_{\mu\nu}$ and $v_{\mu}$. The zeroth order propagator $G^0$
satisfies \be \left[-\partial^2_{0}+\triangle-m^2-2\sum_{q,p\leq
q}b_{q p}\triangle^{q+p}\right] G^0(x,x')=-\delta (x-x'). \ee Hence,
the corresponding Feynman propagator is \be G^{0}_F(x,x')=\int
\frac{d^n k}{(2\pi)^n}
\frac{e^{i k(x-x')}}{[-{k_0}^2+\om^2(|\vec{k}|^2)-i\epsilon] }\ee where $k
x=\eta_{\mu\nu}k^{\mu}x^{\nu}$, and \be
\om^2(|\vec{k}|^2)=m^2+|\vec{k}|^2+2\sum_{q,p\leq q}b_{q
p} (-1)^{q+p}\ |\vec{k}|^{2(q+p)}, \ee with
$\vec{k}=(k_1,...,k_{n-1})$.

Then, the first order correction to the propagator  satisfies
\be
\left[-\partial^2_{0}+\triangle-m^2-2\sum_{q,p\leq q}b_{q
p}\triangle^{q+p}\right]
G^1(x,x')=\left[\bar{h}^{\mu\nu}\partial_{\nu}\partial_{\mu}-\frac{\xi}{2}\eta^{\mu\nu}(\partial_{\nu}\partial_{\mu}h)+\frac{h}{2}(\om^2(-\triangle)+\triangle)+\mathcal{K}\right]G^0(x,x'),
\ee and the corresponding solution is \be
\label{GForden1}G^1_F(x,x')=-\int d^n y
G_F^{0}(x,y)\left[\bar{h}^{\mu\nu}(y)\partial_{\nu}\partial_{\mu}-\frac{\xi}{2}\eta^{\mu\nu}
(\partial_{\nu}\partial_{\mu}h(y))+\frac{h(y)}{2}(\om^2(-\triangle)+\triangle))+\mathcal{K}(y)\right]G^0_F(y,x').
 \ee
In what follows, for the sake of convenience,  we work in
Euclidean spacetime with $k_n=-i k^0$ and  $x_n=-i x^0$. The
Euclidean Green's function $G_E$ is given by $G_F=iG_E$.

For our power counting analysis we consider, as an example, only
one contribution to $G^1_E$, which we choose to be  \be
g^1_E(x,x')=\frac{1}{2}\int d^n y h(y) G^0_E(x,y)\om^2(-\triangle)
G^0_E(y,x'). \label{contrib1}\ee One can readily check that this
term, along with the one proportional to $
\bar{h}^{00}(y)\partial_{0}\partial_{0}$ and similar terms within
$\mathcal{K}(y)$, are the most divergent ones. By replacing the
zeroth order solution $G^0_E$ into this equation, introducing the
Fourier transform of $h$, \be h(p)=\int d^n y e^{-ipy} h(y), \ee
and performing some trivial integrations, we find \be
\label{example} g^1_E(x,x')=\frac{1}{2}\int \frac{d^n p}{(2\pi)^n}
e^{ipx}h(p)\int \frac{d^n k}{(2\pi)^n}\frac{
e^{ik(x-x')}\om^2(|\vec{k}|^2)
}{[(k_n+p_n)^2+\om^2(|\vec{k}+\vec{p}|^2)][k_n^2+\om^2(|\vec{k}|^2)]}.\ee

In order to derive the superficial degree of divergence of each
adiabatic order we start by expanding the integrand in
Eq.(\ref{example}) in powers of $p_i$ $(i=1...n)$,
\begin{subequations}
\begin{align}
g^1_E(x,x')&=\frac{1}{2}\int \frac{d^n p}{(2\pi)^n} e^{ipx}h(p)\int \frac{d^n k}{(2\pi)^n}\frac{ e^{ik(x-x')}\om^2(|\vec{k}|^2) }{[k_n^2+\om^2(|\vec{k}|^2)]^2}\left(\frac{1}{1+\epsilon_p}\right)\\
&=\frac{1}{2}\int \frac{d^n p}{(2\pi)^n} e^{ipx}h(p)\int \frac{d^n
k}{(2\pi)^n}\frac{ e^{ik(x-x')}\om^2(|\vec{k}|^2)
}{[k_n^2+\om^2(|\vec{k}|^2)]^2}\sum_{r=0}^{+\infty}(-\epsilon_p)^r,\label{excomdes}
\end{align} \end{subequations}where \be \epsilon_p=\frac{2k_np_n+p_n^2+\om^2(|\vec{k}+\vec{p}|^2)-\om^2(|\vec{k}|^2)}{k_n^2+\om^2(|\vec{k}|^2)}.\ee

The expectation value $\langle \phi^2 \rangle$ is given by the
coincidence limit of $\mathrm{Re} G_E(=\mathrm{Im} G_F)$. In such
limit, $g^1_E$ can be decomposed as
\begin{subequations}\label{exD}
\begin{align}
g^1_E(x,x)&=\frac{1}{2}\int \frac{d^n p}{(2\pi)^n} e^{ipx}h(p)\mathcal{I}(p),\\
\mathcal{I}(p)&=\int \frac{d^{n-1} k}{(2\pi)^{n-1}} \om^2(|\vec{k}|^2)I(p,|\vec{k}|), \label{excoI}\\
I(p,|\vec{k}|)&=\int \frac{d
k_n}{2\pi}\sum_{r=0}^{+\infty}\frac{(-\epsilon_p)^r}{[k_n^2+\om^2(|\vec{k}|^2)]^2}.
\label{exIch}
\end{align} \end{subequations}
Notice that, due to the remnant symmetries, the terms with odd
powers of $p_n$ or $|\vec{p}|$ do not contribute, so there will be
no odd adiabatic order contributions.

From Eq.(\ref{exD}) it is possible to derive the superficial
degree of divergence of $\mathcal{I}(p)$ for each term with a
given power of $p_i$. However, it is more instructive to compute
explicitly the integral in $k_n$. Moreover, in order to appreciate better
the difference between the cases in which the background fields
depend or not on $\vec{x}$, we analyze them separately.

For background fields that depend only on $x_n$, we replace in Eq.(\ref{exD}) the n-dimensional Fourier
transform $h(p)$ by
$h(p)=(2\pi)^{n-1}\delta^{n-1}(\vec{p})h(p_n)$. Performing the
integral in $k_n$ of Eq.(\ref{exIch}) it is easy to see that
\be\label{r1}
I(p,|\vec{k}|)=\sum_{i=0}^{+\infty}\frac{\alpha_{i}}{\om^3(|\vec{k}|^2)}\left(\frac{p_n}{\om(|\vec{k}|^2)}\right)^{2i},\ee
where $\alpha_{i}$ is independent on $\vec{k}$. Therefore,
substituting Eq.(\ref{r1}) into Eq.(\ref{excoI}), by means of
power counting, we find that when the MDR is such that
$\om(|\vec{k}|^2)\sim|\vec{k}|^s$ for large values of $|\vec{k}|$,
convergence of the $2i-$adiabatic order of $\mathcal{I}(p)$ is
guaranteed if $-s(2i+3)+2s+n-2<-1$. That is, given the highest
power $s$ of $|\vec{k}|$ in the MDR and the spacetime dimension
$n$, we expect that the  maximum  adiabatic order of $\langle
\phi^2 \rangle$ that contains divergences will be given by
\be\label{icasoh} 2i_{max}^h=2\, \mathrm{int}\left(\frac{n-1}{2
s}-\frac{1}{2}\right).\ee This result coincides with the one given
in Eq.(\ref{jmaxhomog}). It is worth to remark that the same
result can be obtained using the weighted power counting analysis
described in Ref.\cite{Anselmi}.

For background fields that depend only on  $\vec x$, we introduce the
spatial Fourier transform $h(\vec{p})$ related to $h(p)$ by
$h(p)=(2\pi)\delta(p_n)h(\vec{p})$. In this case, performing the
integration in $k_n$ we obtain \be
I(p,|\vec{k}|)=\sum_{r=0}^{+\infty}
\frac{\beta_r}{\om^3(|\vec{k}|^2)}
\left(\frac{\om^2(|\vec{k}+\vec{p}|^2)}{\om^2(|\vec{k}|^2)}-1\right)^r,\ee
where $\beta_r$ does not depend on $\vec{k}$.  This equation
should be compared with Eq.(\ref{r1}).  In this case, the odd
powers in $|\vec{p}|$ vanish once the angular integrations in
Eq.(\ref{excoI}) are carried out. For large values of $\vec{k}$,
$\om^2(|\vec{k}+\vec{p}|^2)-\om^2(|\vec{k}|^2)\sim
|\vec{k}|^{2s-2} (\vec{p}\cdot \vec{k})$, then power counting
indicates that  the $2i-$adiabatic order will be finite if
$2i>n-1-s$. Hence, in this case, we expect that divergences will
appear up to the $2i_{max}^g-$adiabatic order of  $\langle \phi^2
\rangle$, where \be\label{imaxgen} 2i_{max}^g=2\,
\mathrm{int}\left(\frac{n-1-s}{2}\right).\ee Here the superscript
$g$ stands for {\it general}, since it can be easily
shown that this is also the superficial
degree of divergence for arbitrary backgrounds (i.e. $h_{\mu\nu}
=h_{\mu\nu}(\vec x,x_n))$, $v_{\mu}=v_{\mu}(\vec x,x_n))$. Note that while
the values of $2i_{max}^g$ given in this equation coincide with
those obtained from Eq.(\ref{icasoh}) for $n\leq 4$, it is
generally larger for higher number of dimensions. For example, for
$n=5$ and $s=2$ we have $2i_{max}^g=2$ while $2i_{max}^h=0$.

We have performed our power counting analysis for the particular
contribution to $G_E^1$ given in Eq.(\ref{contrib1}), and one may
wonder if other contributions to the propagator could cancel the
divergences in $g_E^1$. In order to check explicitly that this is
not the case,  let us consider a MDR of the form
$\om^2(|\vec{k}|^2)=m^2+|\vec{k}|^2+2 b_{11} |\vec{k}|^{4}$ and
compute the operator $\mathcal{K}$ defined in Eq.(\ref{oplin}).
For background fields that depend only on $\vec{x}$, after
performing some integrations in Eq.(\ref{GForden1}), we arrive at
\begin{eqnarray}
\langle \phi^2 \rangle=\mathrm{Re} G_E^1(x,x)&=&-\frac{1}{2}\int \frac{d^{n-1}p}{(2\pi)^{n-1}}e^{i\vec{p}\cdot\vec{x}}\int \frac{d^{n-1}k}{(2\pi)^{n-1}}\left\{\frac{\bar{h}_{00}}{\om(|\vec{k}|^2)+\om(|\vec{k}+\vec{p}|^2)}\right.\nonumber\\
&&\left.+\frac{h(\vec{p})m^2+\xi p^2h(\vec{p})-2\sum_{i,j=1}^{n-1}\bar{h}^{ij}k_ik_j+ f_k(\vec{p})}{2\om(|\vec{k}|^2)\om(|\vec{k}+\vec{p}|^2)[\om(|\vec{k}|^2)+\om(|\vec{k}+\vec{p}|^2)]}\right\},\label{Rege}\end{eqnarray}
where, up to second order of $p_i$,
\be
f_k(\vec{p})= 2 b_{11}\left\{h(\vec{p})|\vec{k}|^4-h_{00}(\vec{p})\left[|\vec{p}|^2|\vec{k}|^2-2 (\vec{p}\cdot\vec{k})^2\right]-2\sum_{i,j=1}^{n-1}{h}^{ij}k_ik_j\left[|\vec{p}|^2+2|\vec{k}|^2+2 \vec{p}\cdot\vec{k}\right] \right\}.
\ee Then, we expand the terms between brackets in Eq.(\ref{Rege}) up to second order in $p_i$ to obtain an integral expression for $\langle \phi^2 \rangle^{(2)}$.
By using dimensional regularization, we perform integrations by parts and discard surface terms to express all the integrals appearing in $\langle \phi^2 \rangle^{(2)}$ in terms of only two of them. In this way, we obtain:
\be\label{phicuadstat}
\langle \phi^2 \rangle^{(2)}=-\frac{\Omega_{n-1}}{8(2\pi)^{n-1}}\left\{I_3\left(\xi-\frac{1}{6}\right) R^{1}+\tilde{I}_3\left[\frac{R^{1}}{6}+R_{00}^{1}\frac{3n-7}{6(n-1)}\right]\right\},
\ee where the factor $\Omega_{n-1}= 2\pi^{(n-1)/2}/\Gamma[(n-1)/2]$ comes from the angular
integration, $R^{1}=-\Delta h/2$,  $R^{1}_{00}=-\Delta h_{00}/2$, and
\begin{subequations}
\begin{align}
I_3&=\int_{0}^{\infty} dx \frac{x^{\frac{(n-3)}{2}}}{\omega^3(x)},\\
\tilde{I}_3&=\int_{0}^{\infty} dx \frac{x^{\frac{(n-1)}{2}}}{\omega^3(x)}\frac{d^2\omega^2(x)}{dx^2}.
\end{align}\end{subequations} Here we see that while the integral $I_3$ converges in $n=5$ dimensions,
there appears a new integral $\tilde{I}_3$ which is proportional
to $b_{11}$ and diverges as $n\to 5$. Moreover, the divergence
proportional to $R_{00}^{1}$ is not purely geometric. Indeed, it can  be covariantly
written in terms of the metric and the aether field since, to linear order in the perturbation fields, $R_{00}^{1}=R_{\mu\nu}u^\mu u^\nu = \nabla_{\mu}a^{\mu}$. On the other
hand, when $\langle \phi^2 \rangle^{(2)}$ is computed for
background fields that depend only on the time coordinate, the
second term in Eq.(\ref{phicuadstat}) does not appear and, in
agreement with Eq.(\ref{icasoh}), this adiabatic order is finite.
However, as cancellations are not to be expected for general
background fields, divergences will generally appear up
to the $2i_{max}^{g}-$adiabatic order given in Eq.(\ref{imaxgen}).

The differences between the two cases we are considering are better
evidenced when the power counting analysis is applied to the stress
tensor of the scalar field. The expectation value of the Euclidean
stress tensor of the scalar field can be expressed as the
coincidence limit of a nonlocal derivative operator applied to the
two-point function $G_E^1(x,x')$. For example, the coincidence limit
of the derivative with respect to $x_n$ and ${x'}_n$ of
$G_E^1(x,x')$ will contribute to the stress tensor. For our analysis
let us consider the following contribution:
\begin{subequations}
\begin{align}
[\partial_n{\partial'}_n g^1_E(x,x')]_{x=x'}&=\frac{1}{2}\int \frac{d^n p}{(2\pi)^n} e^{ipx}h(p)\mathcal{T}(p),\\
\mathcal{T}(p)&=\int \frac{d^{n-1} k}{(2\pi)^{n-1}} \om^2(|\vec{k}|^2)T(p,|\vec{k}|), \label{excoT}\\
T(p,|\vec{k}|)&=\int \frac{d
k_n}{2\pi}\sum_{r=0}^{+\infty}\frac{k_n^2(-\epsilon_p)^r}{[k_n^2+\om^2(|\vec{k}|^2)]^2}.
\end{align} \end{subequations}
When the background fields depend only on $x_n$ we
find that  $T(p,|\vec{k}|)$ can be written as \be
T(p,|\vec{k}|)=\sum_{j=0}^{+\infty}
\frac{\gamma_j}{\om(|\vec{k}|^2)}
\left(\frac{p_n}{\om(|\vec{k}|^2)}\right)^{2j},\ee with $\gamma_j$
k-independent. After substituting this expression into
Eq.(\ref{excoT}) we obtain that divergences are contained in terms
with up to $2j_{max}^h$ powers of $p_n$, where \be 2j_{max}^h=2\,
\mathrm{int}\left(\frac{1}{2}+\frac{n-1}{2 s}\right),\ee in
agreement with Eq.(\ref{jmaxhomog}).

On the other hand, for background fields that depend only on $\vec x$ we
obtain \be T(p,|\vec{k}|)=\sum_{r=0}^{+\infty}
\frac{\zeta_r}{\om(|\vec{k}|^2)}
\left(\frac{\om^2(|\vec{k}+\vec{p}|^2)}{\om^2(|\vec{k}|^2)}-1\right)^{r},\ee
where $\zeta_r$ does not depend on $k$. Recalling that odd powers
in $|\vec{p}|$ do not contribute and using again that
$\om^2(|\vec{k}+\vec{p}|^2)-\om^2(|\vec{k}|^2)\sim
|\vec{k}|^{2s-2} (\vec{p}\cdot \vec{k})$ for large values of
$\vec{k}$, we arrive at \be\label{jmaxgen} 2j_{max}^g=2\,
\mathrm{int}\left(\frac{n-1+s}{2}\right),\ee whence we see that,
contrary to the previous case,  $2j_{max}^g$ increases with $s$.
For instance, for $n=4$ and $s=1,2$ divergences appear up to the
fourth adiabatic order, while for $s=3$ the sixth adiabatic order
is also divergent. Hence, we expect that for $s\geq 3$ the
renormalization of the expectation value of the stress tensor will
require counterterms of adiabatic order higher than four.

Analogously to the case of $\langle \phi^2 \rangle$, for a general
background we do not expect that divergence cancellations occur.
Therefore, we conclude that the subtraction in Eq.(\ref{renor})
should be performed up to the adiabatic orders $2i_{max}^g$ and
$2j_{max}^g$ given in Eqs. (\ref{imaxgen}) and (\ref{jmaxgen}),
respectively. In particular, in order to renormalize the
semiclassical Einstein-Aether equations it will be necessary to
introduce all possible counterterms constructed with $g_{\mu\nu}$
and $u_{\mu}$ up to the $2j_{max}^g-$adiabatic order.

\begin{acknowledgments}

This work has been supported by  Universidad de Buenos Aires, CONICET and ANPCyT.

\end{acknowledgments}

\end{document}